\documentclass[a4paper]{article}
\usepackage{graphicx}
\usepackage{ISCSLP2024}
\usepackage{tabularx}
\usepackage{booktabs}
\usepackage{diagbox}
\usepackage{footnote}
\usepackage[T1]{fontenc}
\usepackage[utf8]{inputenc}
\title{Reject Threshold Adaptation for Open-Set Model Attribution of Deepfake Audio}

\name{Xinrui Yan$^1$, Jiangyan Yi$^2$, Jianhua Tao$^3$$^4$, Yujie Chen$^2$, Hao Gu$^2$, Guanjun Li$^2$, Junzuo Zhou$^2$,\\  
Yong Ren$^2$, Tao Xu$^2$}

\address{
  $^1$Beihang University\\
  $^2$Institute of Automation, Chinese Academy of Sciences\\
  $^3$Department of Automation, Tsinghua University\\
  $^4$Beijing National Research Center for Information Science and Technology, Tsinghua University
}

\email{\texttt{yanxinrui2021@ia.ac.cn, jiangyan.yi@nlpr.ia.ac.cn}}

\begin{document}

\maketitle
\begin{abstract}

Open environment oriented open set model attribution of deepfake audio is an emerging research topic, aiming to identify the generation models of deepfake audio. Most previous work requires manually setting a rejection threshold for unknown classes to compare with predicted probabilities. However, models often overfit training instances and generate overly confident predictions. Moreover, thresholds that effectively distinguish unknown categories in the current dataset may not be suitable for identifying known and unknown categories in another data distribution. To address the issues, we propose a novel framework for open set model attribution of deepfake audio  with \textbf{re}jection \textbf{t}hreshold \textbf{a}daptation (\textbf{ReTA}). Specifically, the reconstruction error learning module trains by combining the representation of system fingerprints with labels corresponding to either the target class or a randomly chosen other class label. This process generates matching and non-matching reconstructed samples, establishing the reconstruction error distributions for each class and laying the foundation for the reject threshold calculation module. The reject threshold calculation module utilizes gaussian probability estimation to fit the distributions of matching and non-matching reconstruction errors. It then computes adaptive reject thresholds for all classes through probability minimization criteria. The experimental results demonstrate the effectiveness of ReTA in improving the open set model attributes of deepfake audio.

\end{abstract}
\noindent\textbf{Index Terms}: model attribution of deepfake audio, system fingerprint recognition, open set recognition

\section{Introduction}
The Text-to-Speech (TTS) generation Application Programming Interface (API) 
\footnote{https://ai.aliyun.com/nls/tts} 
\footnote{https://ai.baidu.com/tech/speech/tts} 
\footnote{https://cloud.tencent.com/product/tts}
provides a method for converting text into natural speech, producing audio that is almost indistinguishable from real human speech \cite{reimao2019dataset}, \cite{yan2022system}, \cite{ma2024leveraging}. However, the misuse of such generation tools or models can lead to the spread of misinformation and violations of suppliers' intellectual property rights \cite{yu2019attributing}, \cite{guan2022you}, among other security issues. Consequently, governments and regulatory bodies are implementing measures \footnote{http://www.cac.gov.cn/2023-07/13/c\textunderscore 1690898327029107.htm} 
\footnote{https://artificialintelligenceact.eu/} to ensure the transparency and traceability of deepfake audio content, including attribution, i.e., identifying the source model of generated contents.

\begin{figure}[t]
  \centering
  \includegraphics[width=\columnwidth]{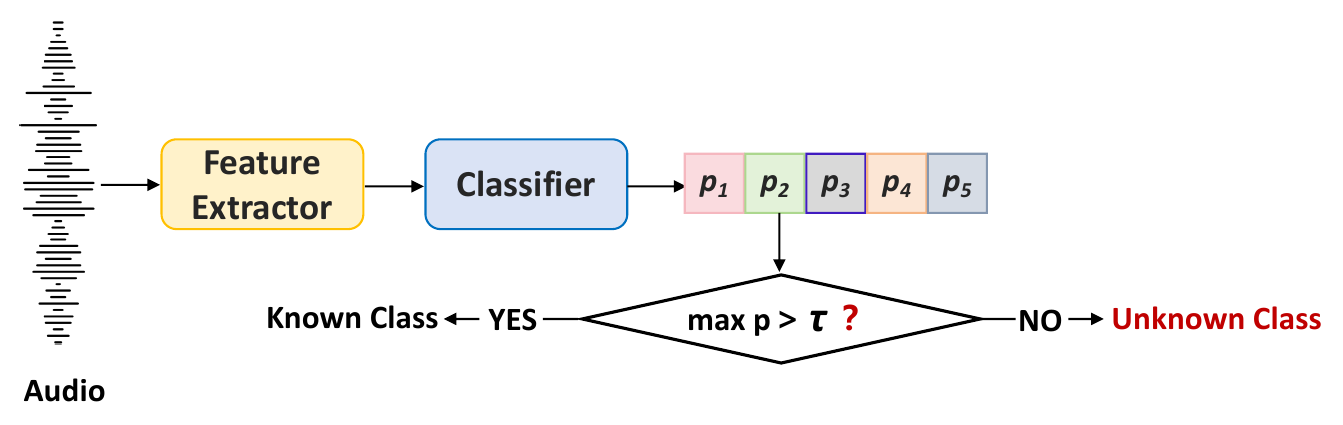}
  \caption{Threshold-based open-set recognition diagram.}
  \label{fig:audio_production}
\end{figure}

\begin{figure}[t]
  \centering
  \includegraphics[width=\columnwidth, height=3cm]{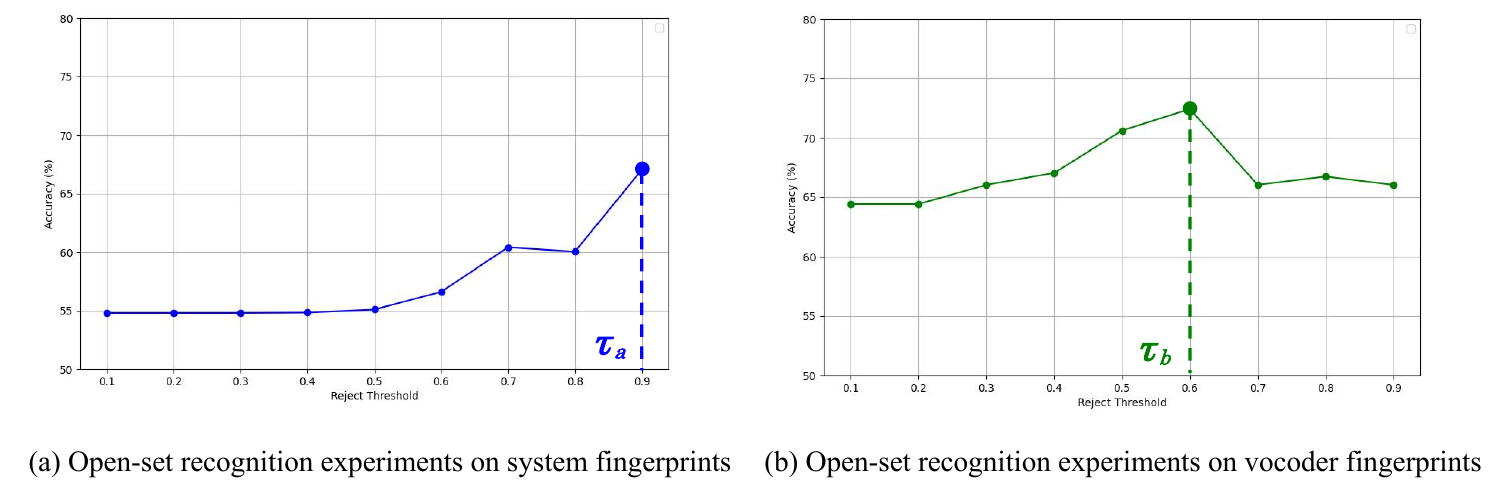}
  \caption{The drawback of manually set threshold-based open-set recognition. The open-set recognition performance in the left and right figures is highly sensitive to the rejection threshold, and the optimal rejection threshold varies for different model attribution tasks.}
  \label{fig:audio_production}
\end{figure}

In real-world open environments, new speech generation models continuously emerge, making it impossible to exhaustively list all possible categories in advance. However, when deploying systems in such environments, it is essential that the recognition system not only accurately classifies known categories but also effectively handles unknown ones. This challenge is commonly referred to as Open Set Recognition (OSR) \cite{geng2020recent}, \cite{kong2021opengan}, \cite{oza2019c2ae}, \cite{park2021divergent}, \cite{liu2022orientational}, \cite{wang2021energy}, \cite{perera2020generative}, \cite{yang2023progressive}, \cite{vaze2021open}. Therefore, for the topic of model attribution of deepfake audio, open set recognition for open environments has important practical significance.

When facing unknown inputs from new categories, an intuitive method to distinguish between known and unknown instances is to manually set a threshold on output probabilities \cite{hendrycks2022baseline}, \cite{li2024open}, as illustrated in Figure 1. \cite{bendale2016towards} first proposed to replace the softmax layer in traditional neural networks with OpenMax. OpenMax calibrates the output probabilities by using the Weber distribution to better handle unknown classes.
\cite{shu2017doc} proposed an alternative to using one-versus-rest units (one-versus-rest units) instead of softmax layers.
\cite{yoshihashi2019classification} utilized a latent representation of the network for reconstruction, leading to robust unknown detection. \cite{oza2019c2ae} proposes the C2AE method, which introduces a new reconstruction task designed to enhance the model's adaptability to open-set scenarios. On track 3 of ADD2023, \cite{lu2023detecting} introduced the kNN distance based on cosine similarity, which separates unknown deepfake audio generation algorithms from known ones based on a threshold criterion. Each of these methods necessitates a manually set threshold to differentiate between known and unknown categories. However, setting and adjusting this threshold is a challenge.

\begin{figure*}[t]
  \centering
  \includegraphics[width=\textwidth]{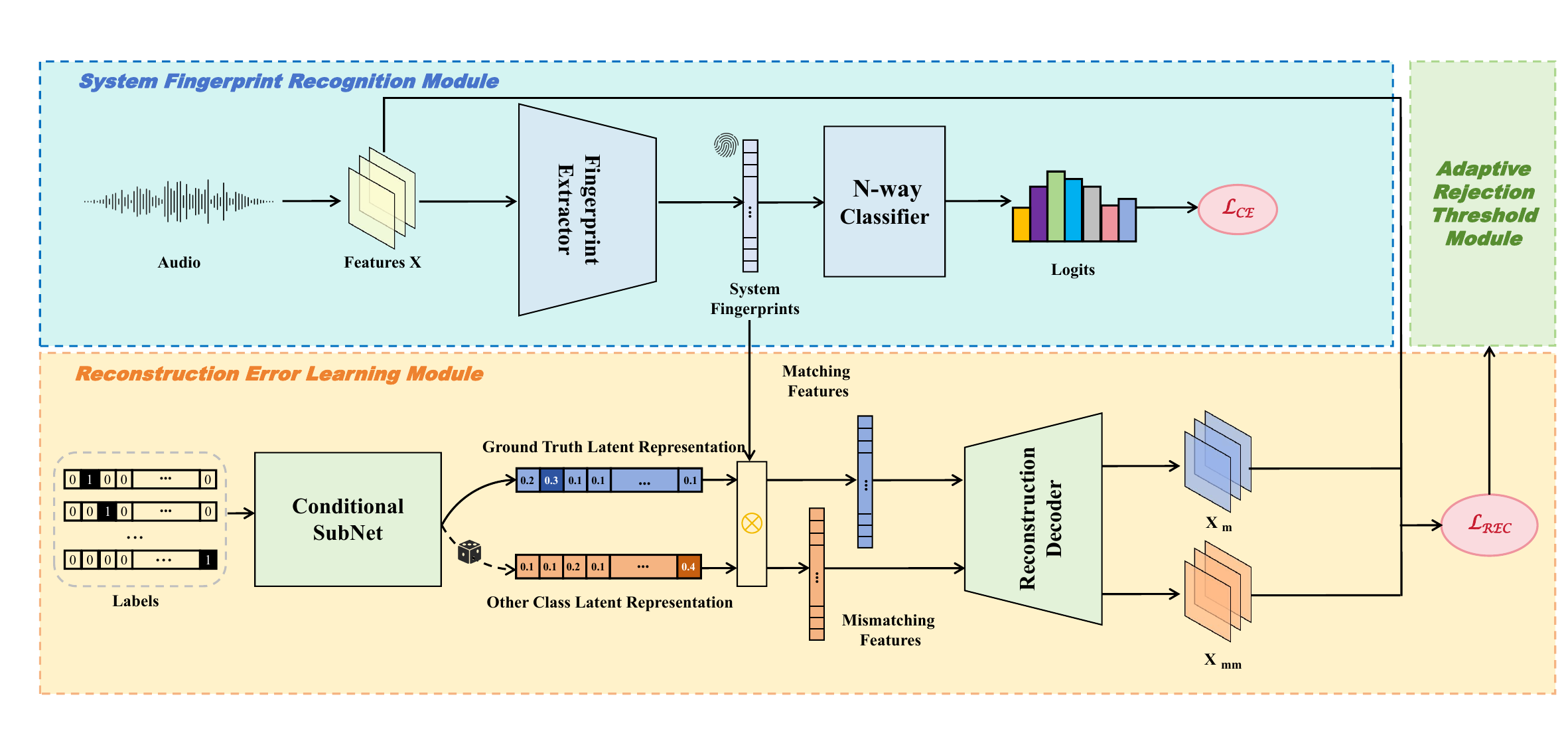}
  \caption{ Overview of the network architecture of ReTA. It consists of 1) system fingerprint recognition module; 2) reconstruction error learning module; 3) adaptive rejection threshold module.}
  \label{fig:audio_production}
\end{figure*}

This method assumes that the model assigns higher probabilities to known categories than to unknown ones. However, deep learning methods often overfit training instances and produce overly confident predictions \cite{Rabiner89-ATO}, \cite{hein2019relu}. Therefore, even for instances from unknown categories, the model may output high probabilities, making the threshold difficult to adjust. Moreover, the composition of speech generation methods is diverse, including different generation tools and algorithms. A threshold-based open set method defines the space below a fixed threshold as the domain of new categories. Due to the different distributions of data from various sources, it is challenging to rely on a single threshold to identify unknown categories with diverse characteristics. The same threshold that effectively separates unknown system fingerprints from known category system fingerprints may not be suitable for identifying unknown vocoder fingerprints \cite{yan2022initial}, as illustrated in Figure 2. Thus, manually setting thresholds poses issues of low accuracy and poor data adaptability.

Motivated by the problems above, we proposed a novel framework  utilizing system fingerprints for open set model attribution of deepfake audio with rejection threshold adaptation (ReTA). The framework aims to calibrate open-set classifiers from two aspects. Specifically, we ensure a significant distinction in the reconstruction error space between known and simulated unknown classes by learning their respective error distributions. This forms the basis for calculating rejection thresholds. Additionally, to better adapt to the open space of unknown classes, ReTA fit adaptive thresholds for the reconstruction errors of each class. Thus, ReTA transform closed-set classifiers into open-set classifiers and adaptively predict class-specific thresholds. Experiments on the open set model attribution dataset SFR validated the effectiveness of our approach in open-set recognition, enhancing the generalizability of threshold-based open-set recognition problems.

Our main contributions are: 
\begin{itemize}
\item We introduced ReTA, the first adaptive threshold framework for open set model attribution of deepfake audio, which enhances the accuracy and data adaptability of the thresholds.

\item We constructed a new system fingerprint encoding-decoding reconstruction network to learn the reconstruction error distributions of known and simulated unknown classes.

\item ReTA effectively improves the capability of open set recognition for model attribution of deepfake audio.
\end{itemize}

This paper is structured as follows: Section II elaborates on the proposed ReTA method in detail. Section III presents comparative experiments conducted on the SFR dataset. In Section IV, we summarize and provide future prospects for the paper.

\section{Proposed Method}

The overall network architecture of ReTA, as shown in Figure 3, consists of three modules: A) System Fingerprint Recognition module; B) Reconstruction Error Learning module; and C) Adaptive Reject Threshold module. The System Fingerprint Recognition module includes 1) fingerprint extraction sub-network; and 2) classification sub-network. The Reconstruction Error Learning module comprises 1) conditional sub-network; and 2) reconstruction decoding sub-network.

\subsection{Framework Architecture}

\subsubsection{System Fingerprint Recognition module}
Firstly, the fingerprint extraction sub-network aims to generate representative high-dimensional fingerprint representations for various classes. The encoder adopts a simple ResNet-18 \cite{he2016deep} architecture with input from low-dimensional LFCC features \cite{sahidullah2015comparison}. Some modifications were made considering the characteristics of deepfake audio, including constraining the activation of the penultimate layer \cite{sun2021react}. The downstream classification sub-network aims to achieve accurate recognition of known classes of system fingerprints. It consists of two convolutional layers that output predicted labels. Finally, the cross-entropy loss $L_{CE}$ \cite{mao2023cross} is computed between the predicted labels and the ground truth labels.

\begin{figure}[t]
  \centering
  \includegraphics[width=\columnwidth]{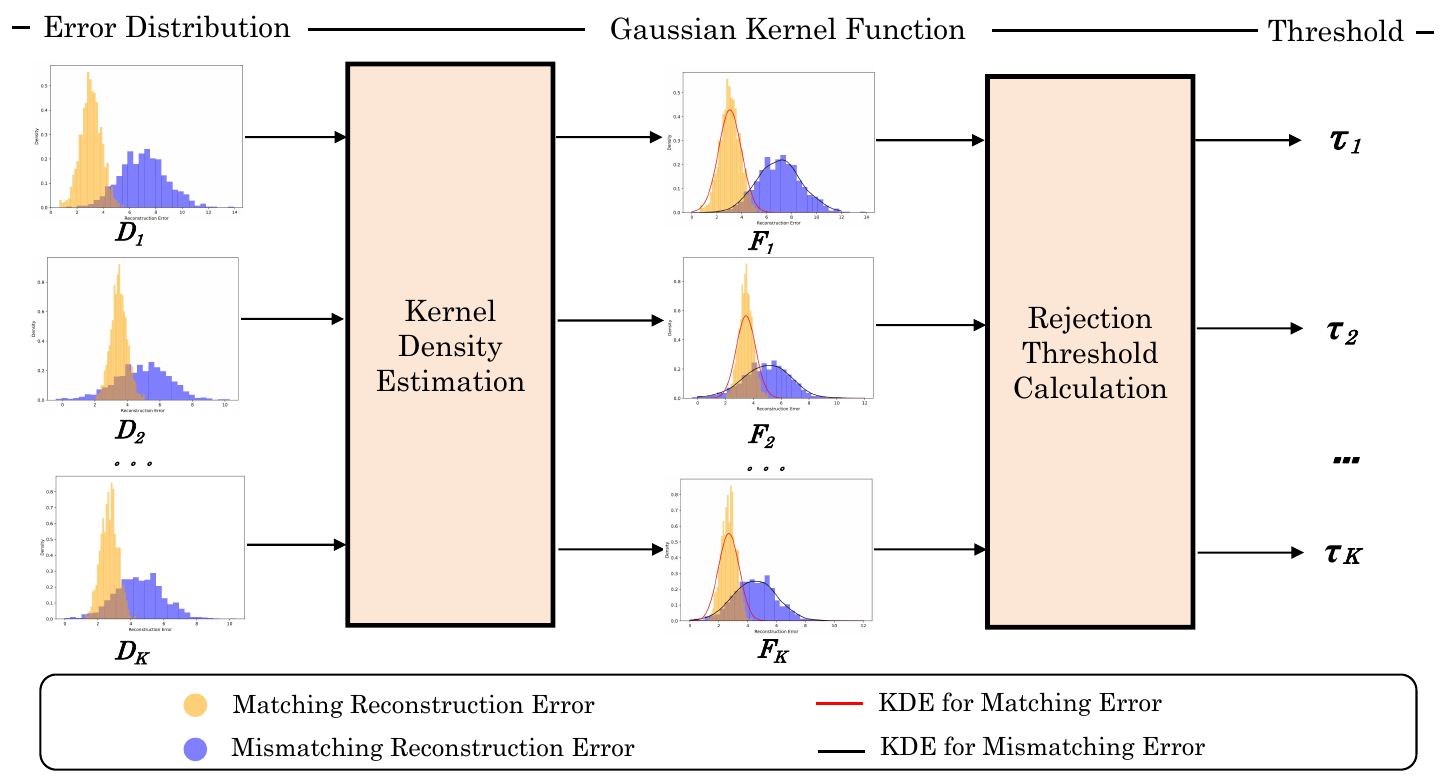}
  \caption{The flowchart of the adaptive reject threshold calculation module.}
  \label{fig:audio_production}
\end{figure}

\subsubsection{Reconstruction Error Learning Module}
The reconstruction decoding sub-network consists of four deconvolution layers, and its input consists of two parts: the fingerprint representation generated by the system fingerprint recognition module and conditional latent features. Specifically, the class labels of known classes are first transformed from the label space to the feature space. These are then input into the conditional sub-network to obtain conditional latent features.

Subsequently, through the reconstruction decoder, the fingerprint representation is randomly multiplied with the corresponding conditional latent features and also with randomly selected conditional latent features from other classes, resulting in matching and non-matching features. Finally, these are used by the reconstruction decoding sub-network to produce matching reconstructed samples $X^m$ and non-matching reconstructed samples $X^{nm}$.

To facilitate better learning of discriminative features between matching and non-matching reconstructed samples in this module, we have also designed a reconstruction error function:
\begin{equation}
L_{\text{REC}} = \frac{1}{N} \sum_{i=1}^{N} \left(\|X_{i} - X_{i}^m\|_2^2 + \|X_{i}' - X_{i}^{nm}\|_2^2 \right)
\end{equation}

where \( X_{i} \) is the input sample,
\(  X_{i}^m \) is the matching reconstructed sample,
\( X_{i}' \) is a randomly sampled sample from other target classes,
\(  X_{i}^{nm}\) is the non-matching reconstructed sample. 
\( N \) represents the total number of known classes.

\subsubsection{Rejection Threshold Calculation Module}

The reject threshold calculation module aims to obtain optimal reject thresholds for each class, as illustrated in Figure 4. This module comprises two operations: Gaussian probability estimation and probability minimization criterion. After computing the matching and nonmatching reconstruction errors for each training sample, the module partitions the set of matching and nonmatching reconstruction errors into N subsets based on class labels. Subsequently, it fits gaussian kernel function with kernel density estimation (KDE) \cite{wkeglarczyk2018kernel}. Finally, optimal reject thresholds are determined using the probability minimization criterion.

\subsection{Training}

The system fingerprint recognition module and the reconstruction error learning module use stochastic gradient descent (SGD) to iteratively update parameters. The overall loss is determined through a weighted sum of \( L_{CE} \) and \( L_{REC} \), expressed as 

\begin{equation}
L= L_{\text{CE}} + \alpha L_{\text{REC}}
\end{equation}

where \( \alpha \) is a weight hyperparameter. The overall loss encapsulates the contributions of both components, effectively shaping and guiding the network's learning process.

\subsection{Inference}
In the open set recognition stage, the test sample undergoes a series of modules. First, the test sample is input into the fingerprint extraction subnetwork to obtain a fingerprint representation. This representation is then fed into the classification subnetwork to generate the predicted class label. Next, the predicted class label and the fingerprint representation are input into the conditional decoding subnetwork to produce the reconstructed sample. Finally, the reconstruction error of this sample is compared with the optimal reject threshold for the predicted class. If the reconstruction error is less than the reject threshold, the test sample is recognized as the predicted class label. Otherwise, the test sample is rejected as belonging to the unknown class.

\section{Experiences}

\subsection{Experimental Settings}

\begin{table*}[ht]
\renewcommand\arraystretch{1}
\centering
\caption{Numbers of utterances for training, development (dev.) and test sets. The number of sentences for clean set and compressed set is comparable. Audio from Tencent and iFLYTEK only appeared as Unknown class during open set recognition.}
\resizebox{\linewidth}{!}{
\begin{tabular}{cccccccccc}
\hline
          Dataset  & Aispeech         & Alibaba Cloud 
             & Baidu Ai Cloud & Databaker   & Sogou & Tencent &iFLYTEK & Real  & Total \\ \hline
{Training} 

       & 26,026   & 31,062  & 7,014          & 10,020      & 10,009   & -- & --     &12,800  & 96,931\\ 
 {Dev.} 
 
    & 8,007   & 10,020  & 2,004          & 2,004      & 4,004     & -- & --   &4,800   & 30,839 \\ 
{Test} 

    & 11,011   & 7,014  & 2,004          & 4,008      & 2,002     & 11,251 & 11,904    &4,800    & 53,994 \\  
{Total} 
& 45,044   &48,096  & 11,022          & 16,032      & 16,015    & 11,251 &11,904    &22,400   & 181,764 \\ \hline
\end{tabular}}
\end{table*}

\begin{table*}[th]
\caption{Performance (\%) comparison on SFR\_clean and SFR\_compressed datasets. The larger the value of each evaluation metric, the better the performance. }
\begin{tabularx}{\textwidth}{@{} l *{9}{>{\centering\arraybackslash}X} @{}}
\toprule
& \multicolumn{3}{c}{SFR\_clean} & \multicolumn{3}{c}{SFR\_compressed} & \multicolumn{3}{c}{Average} \\
\cmidrule(lr){2-4} \cmidrule(lr){5-7} \cmidrule(lr){8-10}
& F$_1$-score & Total Accuracy & ID Accuracy & F$_1$-score & Total Accuracy & ID Accuracy & F$_1$-score & Total Accuracy & ID Accuracy \\
\midrule
Softmax & 72.82 & 60.45 & \textbf{97.82} & 54.87 & 51.92 & 79.71 & 63.85 & 56.19 & 88.77 \\
OpenMax & 88.06 & 67.04 & 97.57 & 58.82 & 57.73 & 78.62 & 73.44 & 62.39 & 88.10 \\
CROSR & 88.94 & 69.23 & 96.47 & 62.38 & 66.59 & 79.95 & 75.66 & 67.91 & 88.21 \\
ReTA (Ours) & \textbf{89.35} & \textbf{71.25} & 96.84 & \textbf{65.34} & \textbf{68.93} & \textbf{81.55} & \textbf{77.35} & \textbf{70.09} & \textbf{89.20} \\
\bottomrule
\end{tabularx}
\label{tab:results}
\end{table*}

ReTA was validated on the SFR dataset. \cite{yan2022system} constructed the first audio dataset for system fingerprint recognition. The audio in the SFR dataset was generated by models from mainstream TTS systems of Chinese vendors, including both clear and compressed sets. The generated speech comes from seven mainstream open source vendors including Aispeech, Alibaba Cloud, Databaker, Sogou, Baidu Ai Cloud, Tencent, and iFLYTEK. A summary of the statistics of Deepfake audio in the SFR dataset is shown in Table 1.

Parameters are randomly initialized. The training was conducted with the Adam optimizer to accelerate optimization by applying adaptive learning rate. The initial learning rate is set to 0.001 for Adam, with linear learning rate decay. These models are trained with a mini-batch size of 128, for 100 epochs.

\subsection{Evaluation Metrics}
To quantitatively evaluate the discriminative performance of our method between known and unknown classes, we introduce $F_1$-score \cite{shmueli2023multi}, Total Accuracy and ID Accuracy to assess various OSR methods. 

$F_1$-score (Macro-average): Used for open-set recognition to balance precision and recall, providing an overall classification indication that includes additional unknown classes. The formulas for calculation are as follows:

\begin{equation}
\text{Precision}_i = \frac{TP_i}{TP_i + FP_i}
\end{equation}

\begin{equation}
\text{Recall}_i = \frac{TP_i}{TP_i + FN_i}
\end{equation}

\begin{equation}
F_{1_i} = 2 \cdot \frac{\text{Precision}_i \cdot \text{Recall}_i}{\text{Precision}_i + \text{Recall}_i}
\end{equation}

where $TP_i$ is the True Positives of class $i$, $FP_i$ is the False Positives of class $i$, and $FN_i$ is the False Negatives of class $i$. Then, calculate the Macro-averaged $F1$-score, which is the average of $F1$-scores for all classes $i$:

\begin{equation}
\text{Macro-averaged } F1\text{-score} = \frac{1}{N} \sum_{i=1}^{N} F_{1_i}
\end{equation}

Total Accuracy: Evaluates the model's ability to correctly classify instances across all classes. It considers correct classifications of known classes as well as correct rejection of unknown classes.

ID Accuracy (Closed-set Accuracy): Measures whether the OSR method can maintain robust classification ability for known classes while rejecting unknown classes.

\subsection{Performance of Model Attribution of Deepfake Audio.}
Our open set recognition performance is compared on the SFR dataset: the test set includes 6 known classes and 2 unknown classes. In Table 2, we compare ReTA with other methods based on manually setting thresholds, including SoftMax (with threshold) \cite{yan2022system}, OpenMax \cite{bendale2016towards}, and the CROSR \cite{yoshihashi2019classification} method. For a fair comparison, all methods use the same LFCC input features and ResNet network structure. Larger values of all evaluation metrics indicate better performance.

Regarding the open set model attribution of deepfake audio, our approach overall achieves the best average performance. Specifically, our method achieves the highest F$_1$-score and total accuracy on both datasets, demonstrating its effectiveness in improving OSR performance. The ultimate goal of open set recognition is not only unknown detection but also maintaining high recognition ID accuracy for known class samples. Our method ensures high sensitivity in identifying unknown classes and robust classification of known classes in compressed datasets, demonstrating strong generalization in compressed environments.

\section{Conclusions}
Manually preset rejection thresholds for model attribution of deepfake audio pose challenges to both the performance of open-set recognition and data adaptability. To address this challenge, we propose a novel adaptive threshold framework named ReTA. ReTA employs a reconstruction error strategy to simulate the unique characteristics of unknown class samples. Utilizing kernel density estimation, ReTA fits Gaussian kernel functions and ultimately computes the optimal reject thresholds for each class through probability minimization criteria. Results obtained on open set model atribution of deepfake
audio demonstrate the effectiveness of the proposed approach. In real-world open scenarios, unknown targets may belong to different classes. Simply rejecting unknown targets as a single ``Unknown'' class is insufficient; there is a need to further distinguish their specific attributes. Therefore, in the future, we will focus on improving this limitation.

\section{Acknowledgments}
This work is supported by the National Natural Science Foundation of China (NSFC) (No. 62322120, No.U21B2010, No. 62306316, No. 62206278).

\bibliographystyle{IEEEtran}

\bibliography{mybib}


\end{document}